# DRL-Based Phase Optimization for O-RIS in Dual-Hop Hard-Switching FSO/RIS-aided RF and UWOC Systems

Aboozar Heydaribeni
Department of Electrical Engineering
Amirkabir University of Technology
(Tehran Polytechnic)
Tehran, Iran
aboozar.hb98@aut.ac.ir

Hamzeh Beyranvand
Department of Electrical Engineering
Amirkabir University of Technology
(Tehran Polytechnic)
Tehran, Iran
beyranvand@aut.ac.ir

Sahar Eslami
Department of Electrical Engineering
Amirkabir University of Technology
(Tehran Polytechnic)
Tehran, Iran
eslamii@aut.ac.ir

***Abstract*—This paper presents a dual-hop hybrid framework that integrates a free-space optical (FSO)/RIS-aided radio frequency (RF) link operating under a hard-switching protocol as the first hop, and an optical reconfigurable intelligent surface (O-RIS)-assisted underwater wireless optical communication (UWOC) link as the second hop. To capture realistic underwater dynamics, the Oceanic Turbulence Optical Power Spectrum (OTOPS) is employed for accurate turbulence modeling. For efficient O-RIS phase control, deep reinforcement learning (DRL) algorithms, specifically the Deep Deterministic Policy Gradient (DDPG) and Twin Delayed DDPG (TD3), have been developed to optimize the phase shifts of O-RIS elements. Simulation results demonstrate that the proposed system substantially improves outage probability and channel capacity, with TD3 achieving superior robustness and adaptability. These findings highlight the DRL-enabled O-RIS as a promising approach for achieving reliable and high-capacity 6G cross-domain UWOC networks.**

## I. INTRODUCTION

The vision of sixth-generation (6G) networks is to enable seamless and unified connectivity across terrestrial, aerial, and underwater environments, supporting mission-critical applications such as cross-domain sensing, autonomous maritime systems, environmental monitoring, and the Internet of Underwater Things (IoUT) [1]. Achieving such integration requires communication architectures capable of adapting to heterogeneous and time-varying physical conditions. Free-space optical (FSO) links, known for fiber-like data rates and low latency, play a central role in high-capacity backhaul; however, their performance degrades under turbulence, weather fluctuations, and pointing errors, motivating the use of hybrid FSO/radio frequency (RF) systems where the RF path serves as a robust backup [2]. Underwater wireless optical communication (UWOC) has emerged as a key enabler for high-speed underwater networking, yet its performance is fundamentally affected by absorption, scattering, and turbulence-induced fading. The Oceanic Turbulence Optical Power Spectrum (OTOPS) offers a physics-based model that captures these impairments as functions of depth, temperature, and salinity, enabling accurate prediction of channel fluctuations [4]. To boost link reliability and extend coverage, optical reconfigurable intelligent surfaces (O-RIS) have been introduced as low-power passive devices capable of shaping the phase of reflected optical waves. Nevertheless, although the recent study in [6] offers comprehensive model-based, heuristic, and ML-driven optimization strategies for RIS configurations, real-time O-RIS control in UWOC systems still constitutes a high-dimensional and non-convex problem, in which classical optimization methods struggle to effectively adapt to the dynamic behavior of the channel. Deep reinforcement learning (DRL) provides a model-free solution that can learn continuous control policies with low online complexity. Actor–critic algorithms such as Deep Deterministic Policy Gradient (DDPG) and Twin Delayed DDPG (TD3) have shown promise for RIS control and ultra-reliable low-latency communication (URLLC) scenarios [7]. However, existing studies typically focus on isolated subsystems and do not offer an integrated end-to-end (E2E) framework that jointly considers: (i) realistic OTOPS-based turbulence modeling for UWOC channels, (ii) practical hard switching between FSO and RF links, and (iii) continuous-action DRL-driven O-RIS phase optimization under hardware constraints. To address these gaps and further substantiate and elaborate on the relevance to International Conference on Information and Knowledge Technology (IKT) topics such as Artificial Intelligence (AI)-based optimization, 6G communication networks, the IoUT, and cross-domain hybrid connectivity, this paper proposes a unified dual-hop architecture that combines FSO and RIS-aided RF links with O-RIS-assisted UWOC channels. The system employs OTOPS-based modeling, integrates a lightweight hard-switching mechanism, and utilizes DDPG/TD3 agents to adaptively optimize O-RIS phases. Monte-Carlo results verify substantial improvements in outage probability (OP) and channel capacity (CC), demonstrating the potential of the proposed approach as a competitive and future-ready solution for 6G-enabled hybrid FSO/RF–UWOC networks.

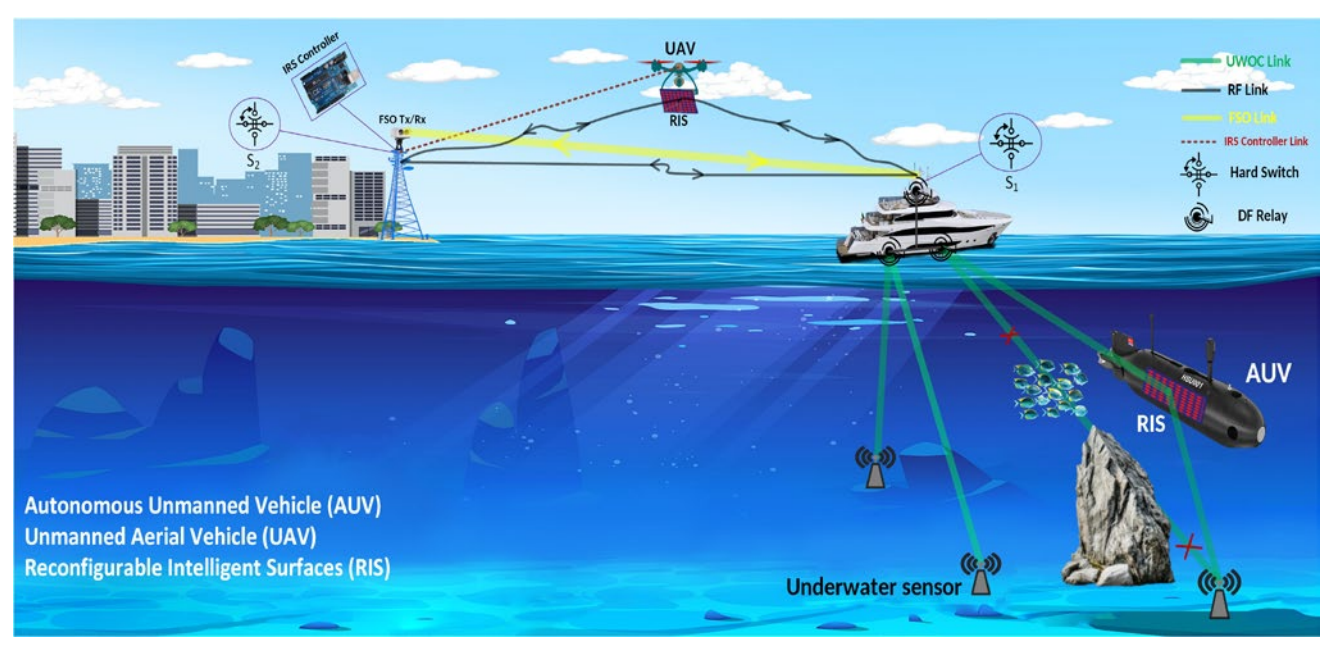

Fig. 1: Schematic of the Proposed Hybrid System Model

## II. SYSTEM MODEL AND PERFORMANCE EVALUATION

In this work, an integrated RF and FSO communication system employing hard switching is investigated between a single-antenna source and a single-antenna destination, as described in [3] and illustrated in Fig. 1. The RF link is assisted by an RIS comprising N passive reflecting elements, which are configured to ensure coherent signal addition from the direct and reflected paths. Under the hard-switching protocol, the FSO link operates as the primary channel when its signal-to-noise ratio (SNR) surpasses a predefined threshold, while the RF link serves as a backup and is activated only when the FSO SNR falls below this threshold. Consequently, the destination receives the signal through either the FSO link or the RF link, but not simultaneously through both.

*A. Free space environment:* In FSO links, the received signal intensity experiences fluctuations due to atmospheric turbulence, which arises from variations in the refractive index caused by temperature and humidity changes. These fluctuations lead to irradiance fading, typically modeled using the Gamma–Gamma (G-G) distribution, which has been shown to be accurate for both moderate and strong turbulence conditions [3]. To simplify the analysis and emphasize the dominant impact of turbulence, perfect transmitter–receiver alignment is assumed, and pointing errors as well as other secondary impairments are neglected [3]. Under this assumption, the instantaneous SNR of the FSO link can be expressed as: $\gamma_{FSO} = \overline{\gamma}_{\mathrm{FSO}}|g_{sd}|^2$, where $\overline{\gamma}_{\mathrm{FSO}}$ denotes the average SNR of the link, and $g_{sd}$ represents the fading coefficient of the optical irradiance. The CDF of the SNR of the FSO channel is given in [3]:

$$F_{\gamma_{FSO}}(\gamma) = \frac{2^{\alpha+\beta-2}}{\pi\Gamma(\alpha)\Gamma(\beta)} G_{1,5}^{4,1}\left(\frac{\alpha^2\beta^2\gamma}{16\overline{\gamma}_{FSO}}\middle| \begin{matrix} 1 \\ \frac{\alpha}{2}, \frac{\alpha+1}{2}, \frac{\beta}{2}, \frac{\beta+1}{2}, 0 \end{matrix}\right) \quad (1)$$

$$\alpha = \left[exp\left(\frac{0.49\sigma_r^2}{\left(1+1.11\sigma_r^{12/5}\right)^{7/6}}\right) - 1\right]^{-1}, \beta = \left[exp\left(\frac{0.51\sigma_r^2}{\left(1+0.69\sigma_r^{12/5}\right)^{5/6}}\right) - 1\right]^{-1}.$$

The key parameters of the G-G model are the scintillation parameters $\alpha$ and $\beta$, which are determined from the Rytov variance $\sigma_{\mathrm{r}}^2 = 1.23 C_n^2 k^{7/6} d_{\mathrm{sd}}^{11/6}$ . The Rytov variance characterizes the strength of turbulence-induced fluctuations and depends on the turbulence strength parameter $C_n^2$, the optical wavenumber $k = 2\pi/\lambda_{\mathrm{FSO}}$, $d_{\mathrm{sd}}$ represents the source-to-destination distance link.

*B- Backup RF Channel Model:* In cases where the quality of the FSO link falls below a predefined SNR threshold, the system switches to the RF link as a backup, as shown in eq. (12) of [3]. In this scenario, the received signal at the destination is composed of the direct transmission from the source as well as the reflected components from the RIS, which can be expressed as $y_{\mathrm{r}} = \sqrt{P}(h_{\mathrm{sd}} + \mathbf{h}_{\mathrm{sr}}^H \Theta \mathbf{h}_{\mathrm{rd}})x + w$. The direct channel $h_{sd}$, the source-RIS channel $h_{sr}$, and the RIS-destination channel $h_{rd}$ are modeled as independent Rayleigh fading processes, where $\lambda_{sd}$, $\lambda_{sr}$, and $\lambda_{rd}$ denote the large-scale fading effects of the corresponding links. The large-scale fading coefficients follow the 3GPP Urban Micro (UMi) path-loss model, which accounts for antenna gains and propagation distance, and are given by [3]:

$$\lambda_l = G_t + G_r - 35.1 - 36.7 log_{10}(d_i), \;\; l = \{sd, sr, rd\} \quad (2)$$

The RIS is represented by a diagonal phase-shift matrix, where each reflecting element controls the amplitude and introduces a controllable phase shift. By synchronizing the phases of the reflected signals with the direct transmission, constructive interference is achieved, thus enhancing the received SNR. Under this ideal condition, the effective SNR depends on the amplitude of the direct channel and the product of the amplitudes of the S-RIS and RIS-D channels.

The effective SNR of the RF link can be expressed as [3]:

$$\gamma_{\mathrm{RF}} = \bar{\gamma}_{\mathrm{RF}}|h_{\mathrm{sd}} + \mathbf{h}_{\mathrm{sr}}^H \Theta \mathbf{h}_{\mathrm{rd}}|^2, \quad (3)$$

The optimal phase of the n-th RIS element is given by:

$$\theta_n^* = \angle h_{sd} - \angle h_{sr_n} - \angle h_{r_n d}, = \theta_{\mathrm{sd}} - (\theta_{\mathrm{sr_n}} + \theta_{\mathrm{r_n d}}). \quad (4)$$

By jointly optimizing all RIS phase shifts, the maximum achievable SNR is:

$$\gamma_{\mathrm{RF}} = \max_{\theta_1,\dots,\theta_N} \bar{\gamma}_{\mathrm{RF}}|h_{\mathrm{sd}} + \mathbf{h}_{\mathrm{sr}}^H \Theta \mathbf{h}_{\mathrm{rd}}|^2, =$$

$$\max_{\theta_1,\dots,\theta_{\mathrm{N}}} \bar{\gamma}_{\mathrm{RF}} \left| |h_{\mathrm{sd}}| e^{j\theta_{\mathrm{sd}}} + \sum_{n=1}^{N} |h_{\mathrm{sr_n}}| . \beta_{\mathrm{n}} . |h_{\mathrm{r_n d}}| e^{j(\theta_{\mathrm{n}} + \theta_{\mathrm{sr_n}} + \theta_{\mathrm{r_n d}})} \right|^2, \quad (5)$$

$$= \bar{\gamma}_{\mathrm{RF}} \left| \alpha_{\mathrm{sd}} + \sum_{n=1}^{N} \alpha_{\mathrm{sr_n}} \beta_{\mathrm{n}} \alpha_{\mathrm{r_n d}} \right|^2, \gamma_{\mathrm{RF}} = \bar{\gamma}_{\mathrm{RF}} \left| \alpha_{\mathrm{sd}} + \sum_{n=1}^{N} \alpha_{\mathrm{sr_n}} \alpha_{\mathrm{r_n d}} \right|^2$$

This expression provides the cumulative distribution function (CDF) of the RF link SNR under optimal RIS phase alignment [3], [9]: $F_{\gamma_{RF}}(\gamma) = 1 - \frac{\Gamma\left(k_o, \frac{\sqrt{\gamma}}{w_o}\right)}{\Gamma(k_o)}$, where $\Gamma(s,x)$ is the upper incomplete Gamma function. The parameters are defined as $k_o = \frac{\left(\sqrt{\lambda_{\mathrm{sd}}\pi} + 2Nkw\right)^2}{4\lambda_{\mathrm{sd}} + 4Nkw^2 - \lambda_{\mathrm{sd}}\pi}$, $w_o = \sqrt{\overline{\gamma}_{\mathrm{RF}}}\left(\frac{4\lambda_{\mathrm{sd}} + 4Nkw^2 - \lambda_{\mathrm{sd}}\pi}{2\left(\sqrt{\lambda_{\mathrm{sd}}\pi} + 2Nkw\right)}\right)$, $k = \frac{\pi^2}{16-\pi^2}$, and $w = \left(4 - \frac{\pi^2}{4}\right)\sqrt{\lambda_{\mathrm{sr}}\lambda_{\mathrm{rd}}}/\pi$. The OP of the mixed FSO/ RIS-assisted RF system operating under a hard-switching scheme is derived as expressed in eq. (14) of [3].

*C. Underwater environment:* The O-RIS-assisted UWOC system comprises *N* reflecting elements, with the transmitted signal x having an average power of $P_t$. The received signal is subject to the combined effects of underwater path loss, turbulence-induced fading, and pointing errors arising from transmitter and O-RIS jitter. Consequently, the received UWOC signal at the DF relay can be expressed as [4]:

$$y = \sum_{n=1}^{N} h_{l_n} h_{t_n} h_{p_n} \rho_n e^{\theta_n} x + n_0, \quad (6)$$

where $h_{t_n}$ denotes the turbulence-induced fading, $h_{p_n}$ represents the fading associated with pointing errors, and $\rho_n \in [0,1]$ is the reflection coefficient of the n-th O-RIS element. Each fading coefficient is expressed as the product of two independent components, i.e., $h_{t_n} = h_{t_{sr_n}} . h_{t_{r_n d}}$, capturing the channel impairments from the source to the O-RIS and from the O-RIS to the destination, respectively. Furthermore, for the ideal case of phase optimization in O-RIS−assisted UWOC links, it is assumed that the optimal phase shift of the n-th element satisfies $\theta_n = -\left(\theta_{sr_n} + \theta_{r_n d}\right)$, ensuring coherent signal combination at the receiver [3]-[4]. Given the negligible physical size and the fact that all of the O-RIS elements lie on a common plane, the pointing error can be assumed identical for all elements, i.e., $h_{p_n} = h_p$ . Likewise, due to the uniform link distances and the small spacing between elements, the path loss is also identical across all elements, $h_{l_n} = h_l$ [4]. In underwater optical communication, path loss mainly results from absorption and scattering caused by water molecules, suspended particles, phytoplankton, and dissolved organic matter. This phenomenon is commonly modeled using Beer–Lambert's law [5] as $h_l = e^{[-c(L_{sr_n} + L_{r_n d})]}$, where $c$ denotes the extinction coefficient ($m^{-1}$), and $L_{sr_n}$ and $L_{r_n d}$ represent the distances from the source to the O-RIS and from the O-RIS to the destination, respectively. The cascaded turbulence fading coefficient, $h_{t_n}$, follows a Gamma–Gamma (G–G) distribution, which accurately captures the combined turbulence effects of both the source-to-O-RIS and O-RIS-to-destination links. Considering an O-RIS with N elements, the sum of independent G–G random variables (RVs) can be approximated by a new G–G RV whose probability density function (PDF) is given in Eq. (8) of [5], and its parameters are defined in Eqs. (8)-(9) of [5]. These parameters are derived from the Rytov variance of each link, assuming that all O-RIS elements experience independent and identically distributed (i.i.d.) fading under plane-wave propagation conditions. The underwater channel turbulence is modeled using the OTOPS [5], which precisely incorporates environmental factors such as temperature and salinity in computing the scintillation index. The required oceanographic parameters are obtained via the Gibbs SeaWater (GSW) Oceanographic Toolbox. The pointing error, denoted by $h_p$, arises from transmitter beam jitter and the surface misalignment of the O-RIS. It can be approximated by $h_p \approx A_0 \exp\left(-\frac{2R^2}{\omega_{z_{eq}}^2}\right)$, where $A_0$ is the fraction of collected power, $\omega_{z_{eq}}$ is the equivalent beam width, and $R$ denotes the radial displacement. Overall, this comprehensive model establishes a rigorous analytical framework for evaluating O-RIS-assisted UWOC systems under realistic impairments, including path loss, turbulence, and pointing errors. Further analytical formulations for the outage probabilities (OPs) of direct and O-RIS-assisted UWOC links can be found in Eqs. (2)-(3) of [4], where the direct-link scenario is also discussed in detail.

*D. System Performance Evaluation through End-to-End (E2E) OP Analysis:* In the E2E OP analysis of the hybrid UWOC−O-RIS and FSO/RIS-aided RF system, assuming statistically i.i.d. fading coefficients across the RIS elements within each link, and statistical independence between the hops, and employing a decode-and-forward (DF) relaying protocol, the cumulative distribution function (CDF) of the overall outage is expressed as [2]: $F_{\gamma_{E2E}}(\gamma) = F_{\gamma_1}(\gamma) + F_{\gamma_2}(\gamma) - F_{\gamma_1}(\gamma) \times F_{\gamma_2}(\gamma)$, where $F_{\gamma_1}(\gamma)$ and $F_{\gamma_2}(\gamma)$ denote the CDFs of the first hop (UWOC with O-RIS assistance) and the second hop (FSO/RF with RIS assistance), respectively [2]. For the UWOC−O-RIS hop, the received signal at the DF relay is given by $y_1 = \sum_{i=1}^{N} \mathcal{H}_i s_1 + n_1$, where $\mathcal{H}_i$ represents the composite i.i.d. channel coefficient, and $N$ denotes the number of O-RIS elements. Similarly, for the FSO/RIS-aided RF hop, the received signal at the destination is modeled as $y_2 = \sum_{k=1}^{M} \mathcal{H}_k s_2 + n_2$. Under the DF protocol, the relay decodes the received signal prior to retransmission; hence, the E2E OP depends on the minimum SNR of both hops and can be written as:

$$OP_{E2E}(\gamma) = OP_1(\gamma) + OP_2(\gamma) - OP_1(\gamma) \times OP_2(\gamma). \quad (7)$$

*E. DRL-Enabled Phase Shift Optimization for O-RIS−Assisted UWOC Systems*:

In this subsection, we establish the DRL-based formulation for practical phase shift optimization in O-RIS-assisted UWOC. The design is cast as a Markov decision process (MDP) defined by the tuple $(\mathcal{S}, \mathcal{A}, \mathcal{R}, \mathcal{P})$, where $\mathcal{S}$ denotes the state space, $\mathcal{A}$ the action space, $\mathcal{R}$ the reward function, and $\mathcal{P}$ the state transition dynamics.

*O-RIS Phase and Amplitude Model*:

Following the practical reflection model in [7] , the O-RIS reflection coefficient is defined as $\Theta = diag\left(\beta_1 e^{j\theta_1}, \dots, \beta_N e^{j\theta_N}\right)$, where $\theta_n \in [-\pi, \pi)$ is the controllable phase of the $n-th$ reflecting element, and the amplitude response depends on the phase as [7]:

$$\mathbf{C_1}: \ \beta_n(\theta_n) = (1 - \beta_{min}) \left(\frac{\sin(\theta_n - \varphi) + 1}{2}\right)^{\alpha} + \beta_{min}. \quad (8)$$

where this nonlinear term reflects the dependence arising from practical hardware limitations of the O-RIS elements.

*State Space*:

The state space is constructed based on the instantaneous effective channel coefficients available to the agent. Since the effective UWOC channel incorporates turbulence-induced fading, pointing errors, path loss, and O-RIS reflection, the state vector at time step $t$ is represented as

$$s_t = [|h(t)|, \angle h(t), |h_{old}(t)|, \angle h_{old}(t)]^T,$$

where $h(t)$ is the current channel realization, and $h_{old}(t)$ is a temporally correlated, outdated estimate modeled with a correlation coefficient $\rho_d$. This representation enables the agent to capture both present and past channel dynamics based on local channel state information (CSI), where the outdated CSI is modeled as follows:

$$h(t + T_d) = \rho_d h(t) + e(t), \quad e(t) \sim \mathcal{CN}\left(0, \sqrt{1 - \rho_d^2 I}\right), \quad (9)$$

where $\rho_d$ is the correlation coefficient and $e(t)$ is a Gaussian innovation process.

*Action Space*:

The action corresponds to the configuration of the O-RIS phase shifts. For an O-RIS consisting of $N$ reflecting elements, the action vector is defined as $a_t = [\theta_1(t), \theta_2(t), \dots, \theta_N(t)]^T$, with each controllable phase constrained to the feasible interval $\theta_n(t) \in [-\pi, \pi) \quad \forall n \in \{1, 2, \dots, N\}$. To reflect hardware limitations, the continuous action space is quantized into $Q = 2^b$ discrete levels after action generation, given by

$$\mathbf{C_2}: \hat{\theta}_n = Q(\theta_n) = \{-\pi, -\pi + \Delta, \dots, -\pi + (Q - 1)\Delta\}, \quad (10)$$

where $\Delta = \frac{2\pi}{2^b}$ , and $b$ denotes the phase resolution bits.

*Reward Function*:

The reward is designed to incentivize reliable communication while penalizing abrupt O-RIS reconfiguration. The reward is designed to capture finite block length (FBL) effects, outage performance, and reconfiguration smoothness. In the FBL regime, the decoding error probability is expressed as [7]:

$$\varepsilon = Q\left(\sqrt{\frac{m}{V(SNR)}}\left(log_2(1 + SNR) - \frac{L}{m}\right)\right), \quad (11)$$

where $m$ is the blocklength, $L$ the number of information bits, and $(V(\cdot))$ the channel dispersion. The corresponding achievable rate is $R = log_2(1 + \tau.SNR) - \sqrt{\frac{V(SNR)}{m}} Q^{-1}(\varepsilon) + O\left(\frac{log(m)}{m}\right)$. The instantaneous reward at step $t$ is expressed as

$$r_t = \alpha_r \cdot \left(1 - 1_{\gamma(t) < \gamma_{th}}\right) + (1 - \alpha_r) \cdot \left(1 - \frac{max\ \{0, \gamma_{th} - \gamma(t)\}}{\gamma_{th}}\right) - \lambda_{reg} |\theta(t) - \theta(t-1)|^2, \quad (12)$$

where $\gamma(t) = \bar{\gamma}|h(t)|^2$ is the instantaneous received SNR, $\gamma_{th}$ is the outage threshold, $\alpha_r \in [0,1]$ balances reliability and throughput maximization, and $\lambda_{reg}$ is the regularization factor penalizing large phase updates. Thus, the optimization at each transmission time (Problem Formulation) reduces to

$$\boldsymbol{P}: (K = 1): \ \max_{\Theta} \ L(\Theta) = V(\Theta) - Q^{-1}\left(\varepsilon^{th}\right) W(\Theta), \quad \textbf{s.t.}: C_1, C_2 \quad (13)$$

*DRL Algorithm Integration*:

Since the action space is continuous, the optimization is addressed using actor–critic based DRL algorithms. In this context, the DDPG algorithm employs deterministic policy gradients to learn the phase adjustment policy; however, it suffers from the problem of Q-value overestimation.

**Algorithm 1** Training and Evaluation of TD3 vs DDPG for UWOC-RIS System

1: **Input:** number of episodes $N_{\text{ep}}$, steps per episode $T$, replay buffer size, mini-batch size $|\mathcal{B}|$, discount $\gamma$, polyak $\tau$, TD3 delay $d$, noise stds and clipping params, quantization bit $b$.
2: **Initialize:**
3: Actor and Critic networks for TD3: $\mu_{\text{TD3}}(\cdot;\xi^{act}_{\text{TD3}})$, $Q^{\text{TD3}}_{1,2}(\cdot;\xi^{crit}_{1,2,\text{TD3}})$ and target copies $\xi^{targ-act}_{\text{TD3}}, \xi^{targ-crit}_{1,2,\text{TD3}}$.
4: Actor and Critic networks for DDPG: $\mu_{\text{DDPG}}(\cdot;\xi^{act}_{\text{DDPG}})$, $Q^{\text{DDPG}}(\cdot;\xi^{crit}_{\text{DDPG}})$ and targets $\xi^{targ-act}_{\text{DDPG}}, \xi^{targ-crit}_{\text{DDPG}}$.
5: Replay buffers $\mathcal{B}_{\text{TD3}}$, $\mathcal{B}_{\text{DDPG}}$.
6: Randomly initialize RIS phase matrix.
7: **for** episode $= 1$ to $N_{\text{ep}}$ **do**
8: Initialize environment state $s \leftarrow s_0$ (collect CSI $\{h_i(t)\}_{i=1}^{K}$)
9: **for** step $t = 1$ to $T$ **do**
10: **Action selection (common):**
11: TD3: $a_{\text{TD3}} \leftarrow \mu_{\text{TD3}}(s;\xi^{act}_{\text{TD3}}) + \kappa_{\text{TD3}}$, where $\kappa_{\text{TD3}} \sim \mathcal{N}(0,\sigma^2_{\text{TD3}})$ (clipped).
12: DDPG: $a_{\text{DDPG}} \leftarrow \mu_{\text{DDPG}}(s;\xi^{act}_{\text{DDPG}}) + \kappa_{\text{DDPG}}$, where $\kappa_{\text{DDPG}} \sim \mathcal{N}(0,\sigma^2_{\text{DDPG}})$ (clipped).
13: No-Learning baseline: random quantized action $a_{\text{rand}}$.
14: Quantize actions if required: $a_q \leftarrow \text{Quantize}(a, b)$.
15: Interact with environment: $(s', r, \text{info}) \leftarrow \text{EnvStep}(s, a_q)$.
16: Store $(s, a_q, r, s')$ in buffers (as applicable).
17: **Sample mini-batch:** $\{(s_j, a_j, r_j, s'_j)\}_{j=1}^{|\mathcal{B}|}$ from each buffer.

---

18: **TD3: Critic update (per training step):**
19: **for** each sample $j$ **do**
20: Compute target action with smoothing noise:
$\tilde{a}'_j \leftarrow \text{clip}\left(\mu_{\text{TD3}}(s'_j;\xi^{targ-act}_{\text{TD3}}) + \epsilon_j,\ -a_{\max}, a_{\max}\right)$
where $\epsilon_j \sim \mathcal{N}(0,\sigma^2_{\text{target}})$ (clipped).
21: Compute target Q-value:
$y^{\text{TD3}}_j = r_j + \gamma \min_{i\in\{1,2\}} Q^{\text{TD3}}_i(s'_j, \tilde{a}'_j;\xi^{targ-crit}_{i,\text{TD3}})$
22: **end for**
23: Update critics $\xi^{crit}_{i,\text{TD3}}$ by minimizing:
$\frac{1}{|\mathcal{B}|}\sum_j \left(Q^{\text{TD3}}_i(s_j, a_j;\xi^{crit}_{i,\text{TD3}}) - y^{\text{TD3}}_j\right)^2, \quad i = 1, 2$

24: **TD3: Delayed actor & target update:**
25: **if** $t \bmod d = 0$ **then**
26: Update actor $\xi^{act}_{\text{TD3}}$ via:
$J_{\text{TD3}} \approx \frac{1}{|\mathcal{B}|}\sum_j Q^{\text{TD3}}_1(s_j, \mu_{\text{TD3}}(s_j;\xi^{act}_{\text{TD3}});\xi^{crit}_{1,\text{TD3}})$
27: Soft-update target networks (Polyak):
$\xi^{targ} \leftarrow \tau\xi + (1-\tau)\xi^{targ}$
28: **end if**

---

29: **DDPG: Critic update (per training step):**
30: **for** each sample $j$ **do**
31: $a'_j \leftarrow \mu_{\text{DDPG}}(s'_j;\xi^{targ-act}_{\text{DDPG}})$
32: $y^{\text{DDPG}}_j = r_j + \gamma Q^{\text{DDPG}}(s'_j, a'_j;\xi^{targ-crit}_{\text{DDPG}})$
33: **end for**
34: Update critic $\xi^{crit}_{\text{DDPG}}$ by minimizing:
$\frac{1}{|\mathcal{B}|}\sum_j \left(Q^{\text{DDPG}}(s_j, a_j;\xi^{crit}_{\text{DDPG}}) - y^{\text{DDPG}}_j\right)^2$

35: **DDPG: Actor update & target update (per training step):**
36: Update actor $\xi^{act}_{\text{DDPG}}$ via:
$J_{\text{DDPG}} \approx \frac{1}{|\mathcal{B}|}\sum_j Q^{\text{DDPG}}(s_j, \mu_{\text{DDPG}}(s_j;\xi^{act}_{\text{DDPG}});\xi^{crit}_{\text{DDPG}})$
37: Soft-update DDPG target networks:
$\xi^{targ} \leftarrow \tau\xi + (1-\tau)\xi^{targ}$

---

38: **Update training stats:** accumulate rewards, store previous action, set $s \leftarrow s'$.
39: **end for**
40: Compute average rewards for this episode.
41: **end for**
42: **Evaluation:** For each average SNR $\bar{\gamma}$:
43: Execute multiple Monte-Carlo trials using deterministic policies with action quantization applied as required.
44: Compute Outage Probability (OP), Average instantaneous SNR, Channel Capacity (CC).
45: **Output:** Trained TD3 and DDPG actors (weights), evaluation curves (OP, SNR, CC).

To overcome this limitation, the TD3 algorithm has been introduced, which incorporates three key mechanisms to enhance learning performance. First, clipped double Q-learning is applied by utilizing two critic networks simultaneously to mitigate overestimation bias. Second, target policy smoothing is achieved by injecting clipped Gaussian noise into the target action. Finally, delayed policy updates are employed to improve the stability of the learning dynamics. Specifically, the target action in TD3 is given by:

$$a' = clip(\mu(s'; \xi^{targ\text{-}act}) + \epsilon,\ a_{min},\ a_{max}), \qquad (14)$$

where $\epsilon \sim \mathcal{N}(0, \sigma^2)$is the smoothing noise clipped within $[-c, c]$, and $a_{min} = -\pi, a_{max} = \pi$, where $\mu(\cdot)$ is the actor network mapping state $s'$ to the target action. The proposed MDP formulation allows the DRL agent to adaptively learn O-RIS phase configurations that minimize OP, improve average SNR, and maximize channel capacity under realistic underwater impairments. Compared with conventional optimization-based techniques, this learning-based framework is computationally efficient and suitable for real-time deployment in IoUT-enabled UWOC scenarios, including autonomous underwater vehicles, industrial inspection, and large-scale environmental monitoring. The design aligns with contemporary trends in O-RIS-assisted short-packet wireless communications [7]-[8], while extending them to the challenging domain of optical underwater environments (see [7] for further details).

## III. SIMULATION RESULTS AND DISCUSSION

The main parameters used in this study are listed in Table 1. For analysis, the RIS is assumed to operate under a fully coherent model in ideal conditions, i.e. $(\theta_n + (\theta_{sr_n} + \theta_{r_n d}) = 0)$and $(\rho_n = 1)$; however, in the DRL evaluation, both the amplitude and phase are not fixed. Fig.2 illustrates the OP of the UWOC channel-assisted O-RIS versus the average SNR under varying temperatures, salinities, and the number of O-RIS elements. The results indicate that a larger number of O-RIS elements or lower water temperature and salinity decrease the OP, thereby enhancing system performance.

Tab. 1. TABLE OF PARAMETERS [2]-[8]

| Parameter | Value | Parameter | Value | Parameter | Value |
|---|---|---|---|---|---|
| $\gamma_{th,FSO}$ | 7, 10 dB | $\gamma_{th,RF}$ | 3 dB | $\gamma_{th,UWOC}$ | 15 dB |
| $\sigma^2_{n,O-RIS}$ | -100 dBm | $\sigma^2_{n,D}$ | -95.5 dBm | $\sigma_\theta$ | 2 mrad |
| $\sigma_{S,F_D}$ | 0.2 | *Water Type* | Jerlov-IB | $\lambda_{UWOC}$ | 532 nm |
| $\sigma_\varphi$ | 1.5 mrad | $\lambda_{FSO}$ | 1550 nm | $f_c$ | 3 GHz |
| $\alpha_{Reward}$ | 0.7 | $b$ | 3 bits | *Replay Buffer Size* | $10^5$ |
| $N_{M.C.}$ | $10^6$ | $W_0$ | 0.01 m | $D_R$ | 0.05 m |
| $\lambda_{RF}$ | 10 cm | $RF_{Gain}$ | 44 dB | $C^2_n$ | $10^{-13}$ m$^{-2/3}$ |
| $x_T$ | $10^{-5}$ K²/s | $H$ | -0.2 $°C.ppt^{-1}$ | $\beta_0$ | 0.72 |
| $SA$ | 5, 20 g/kg | $P$ | 0 dbar | $T$ | 0°, 15°C |

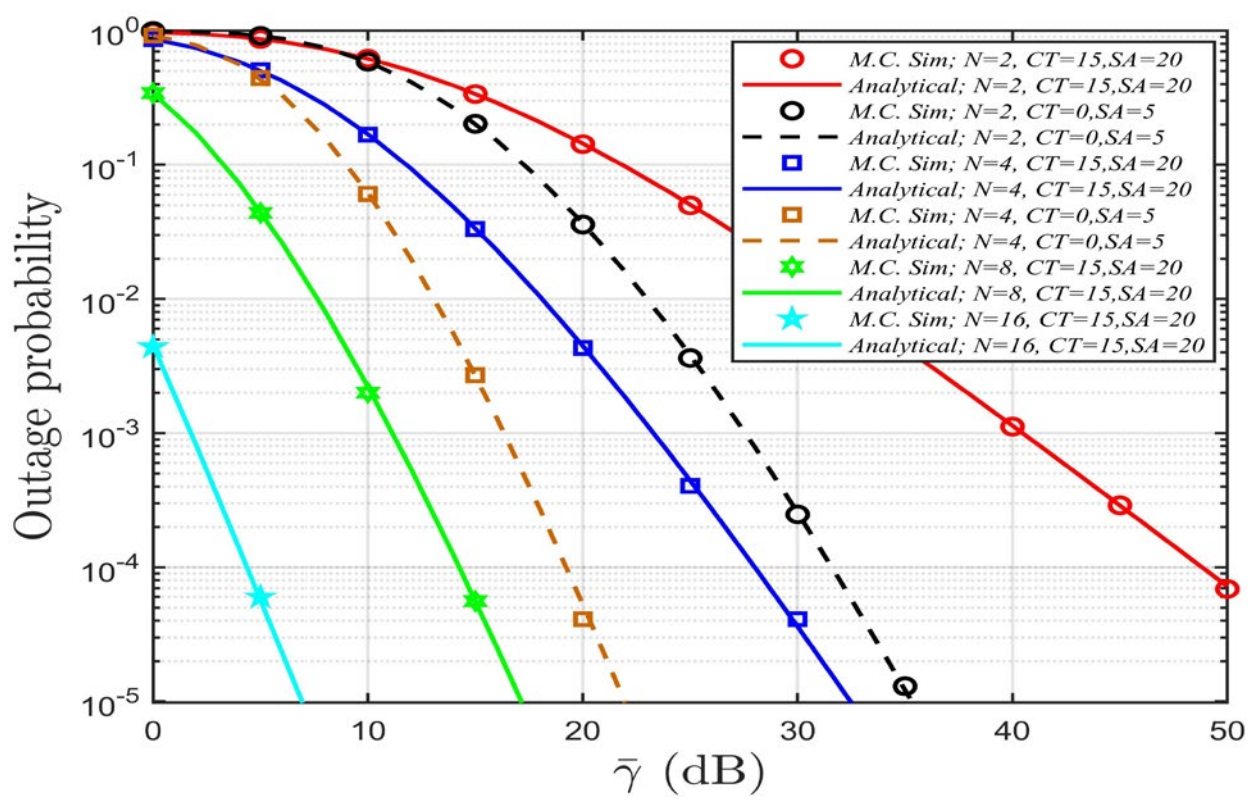

Fig. 2. Outage probability versus $\bar{\gamma}$ under different numbers of O-RIS elements, temperature, and salinity.

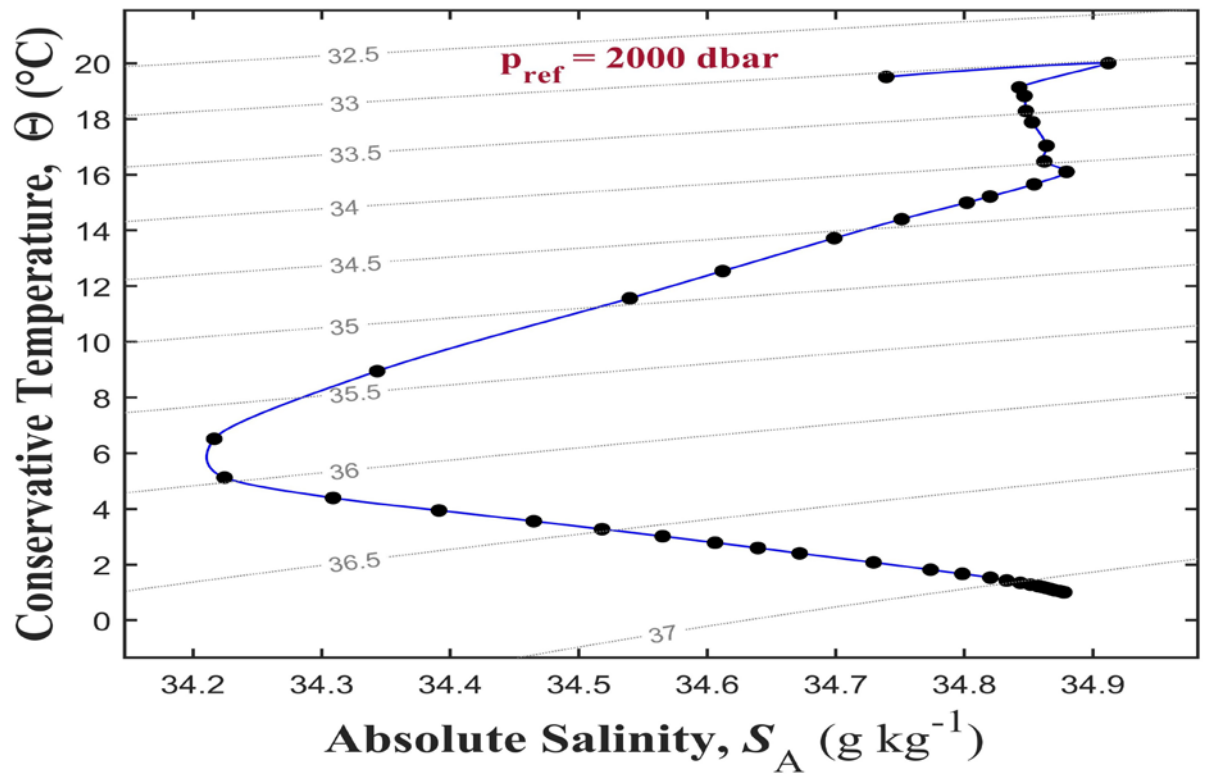

Fig. 3. T–S diagram at 2000 dbar reference pressure.

Fig. 3 illustrates the T–S diagram at 2000 dbar, showing the stability of stratified water masses. Temperature and salinity variations reveal the stratification and stability of different water masses, reflecting the accuracy of the simulations conducted in this study, which were implemented using the GSW MATLAB toolbox. Fig. 4 illustrates the OP of the FSO/RIS-aided RF system versus $\bar{\gamma}_{FSO}$ under hard-switching for different numbers of RIS elements. The use of RIS provides a significant performance improvement compared to the case without RIS, and increasing the number of elements further enhances the performance, particularly in the low-SNR regime where the FSO link is weak. Moreover, a higher switching threshold results in an increased OP, while an increase in the FSO link SNR improves the system performance. The Monte-Carlo simulation results are also shown to be in good agreement with the analytical expressions. Fig.5 illustrates the E2E OP versus the average SNR, for direct UWOC and FSO links with $d_{sd} = 120$m and 1000m, respectively, as well as O-RIS-assisted UWOC and FSO/RIS-aided RF links, where $\bar{\gamma}_{RF} = 38dB$ , $d_{sr} = 50$m, 960m and $d_{rd} = 50$m, 70m . The results show that increasing the number of RIS elements in both links reduces the OP and enhances system performance. As illustrated in Fig. 6, the DRL-based algorithms, particularly TD3, achieve a significantly higher average reward than DDPG and the non-learning scheme, demonstrating enhanced decision-making efficiency of the UWOC system, with the training and evaluation procedures summarized in *Algorithm 1* to ensure fair comparison. In Fig. 7, the OP versus the average SNR is presented for $N_{O-RIS} = 32$ and $L_{sr} = L_{rd} = 40m$ . The results demonstrate that both TD3 and DDPG outperform the non-learning approach by substantially reducing the outage probability. Moreover, the TD3 algorithm exhibits a steeper decline in the outage curve compared to DDPG, thereby offering improved reliability and robustness of the UWOC link. Finally, Fig. 8 depicts the channel capacity under IM/DD detection scheme with the same parameters, where DRL notably enhances performance. Among the evaluated methods, TD3 consistently achieves the highest channel capacity across different SNR regimes, further highlighting its superiority over DDPG.

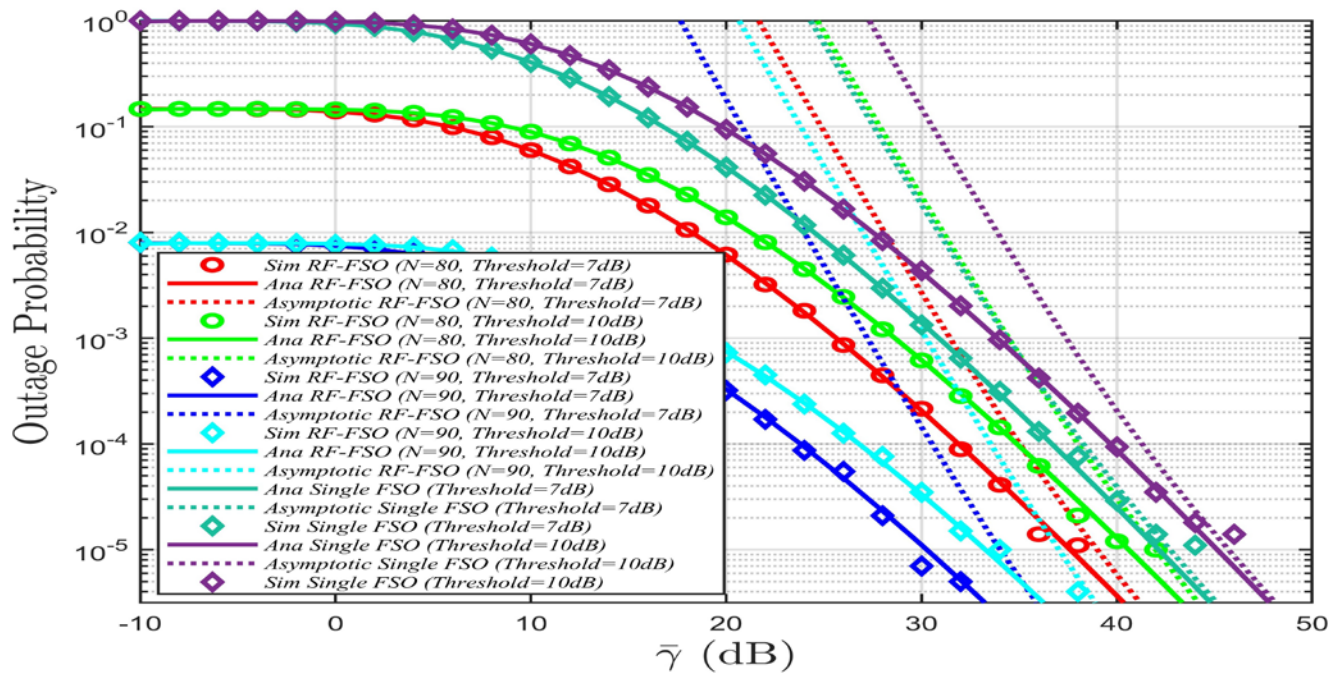

Fig. 4. Outage probability versus $\bar{\gamma}_{FSO}$ with $\gamma_{th,RF} = 3$ dB, $\bar{\gamma}_{RF} = 24.5$dB, $d_{sr} = 450m, d_{rd} = 60m, d_{sd} = 500m$, for different numbers of RIS elements and FSO link thresholds ($\gamma_{th,FSO}$).

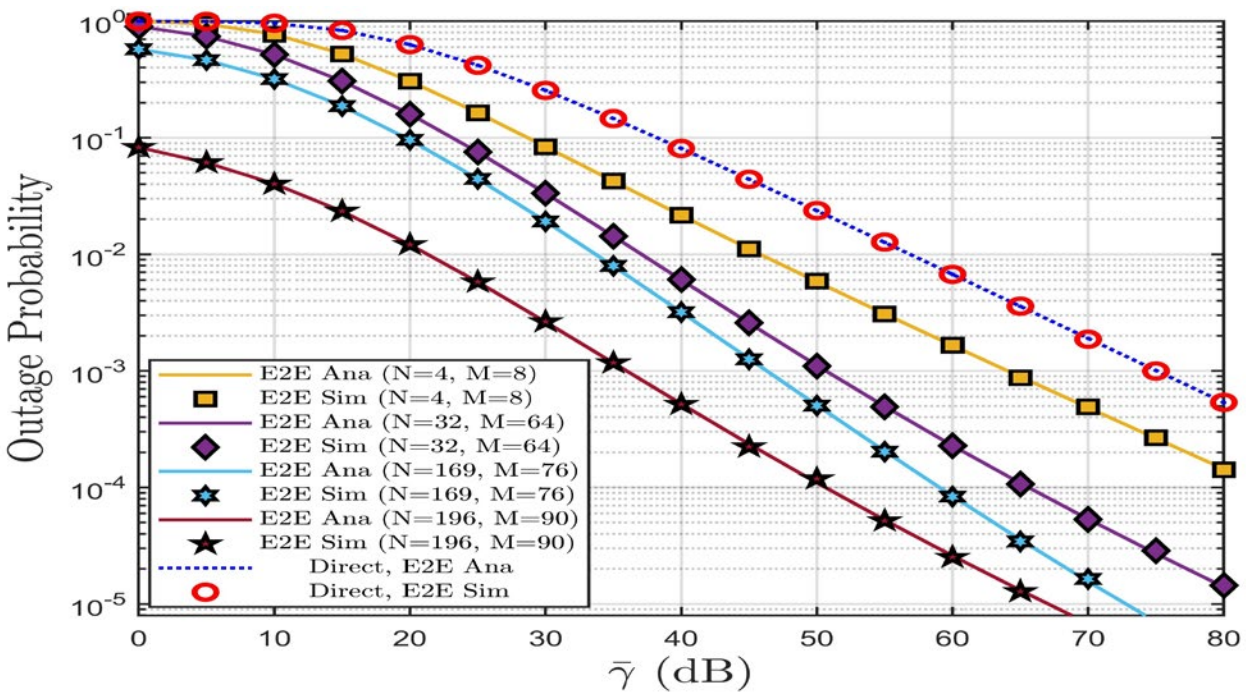

Fig. 5. E2E outage probability versus $\bar{\gamma}$ for direct UWOC and FSO links, and O-RIS-assisted UWOC and FSO/RIS-aided RF links, with different numbers of RIS elements in both links.

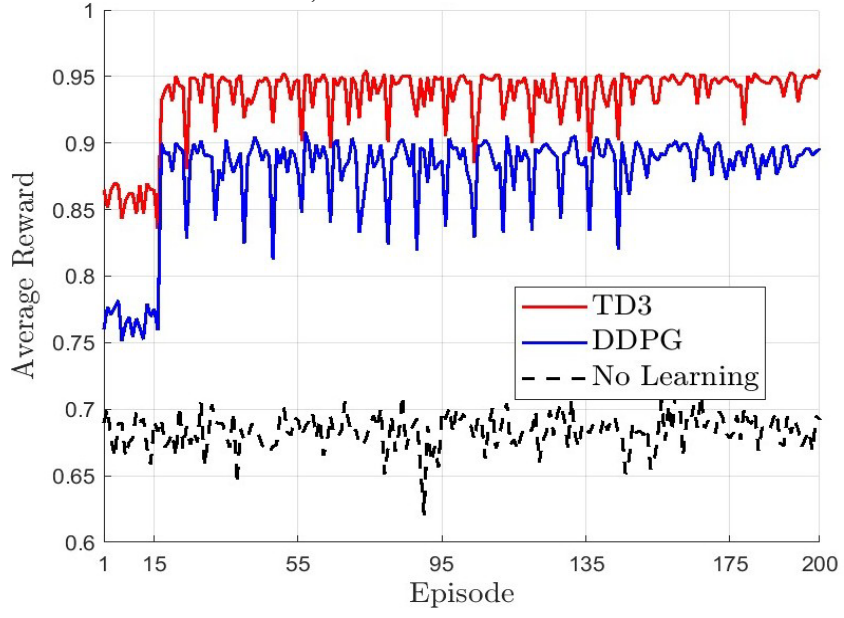

Fig. 6. Average reward of the UWOC−O-RIS channel under DRL-based TD3, DDPG, and no learning.

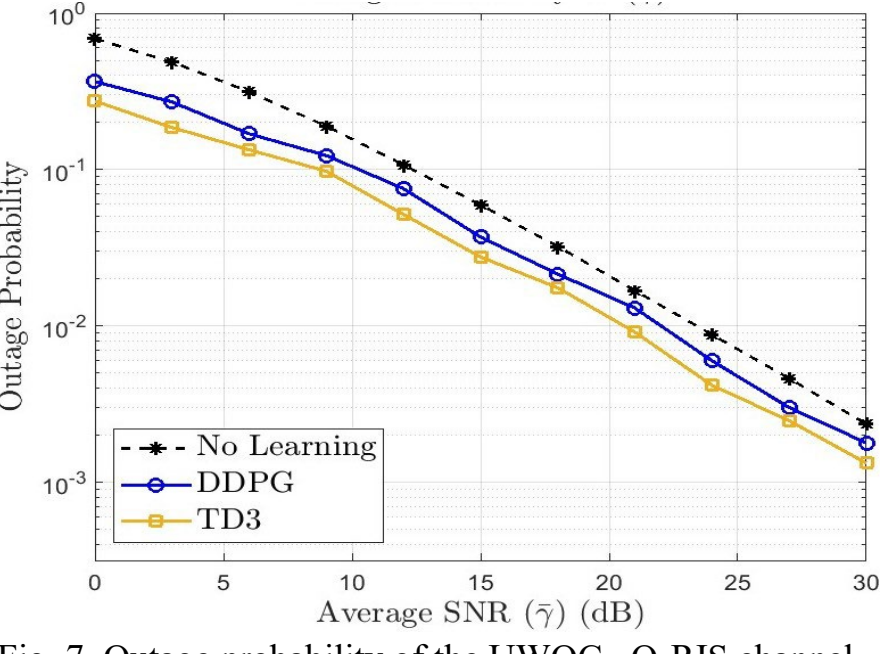

Fig. 7. Outage probability of the UWOC−O-RIS channel under DRL-based TD3, DDPG, and no learning.

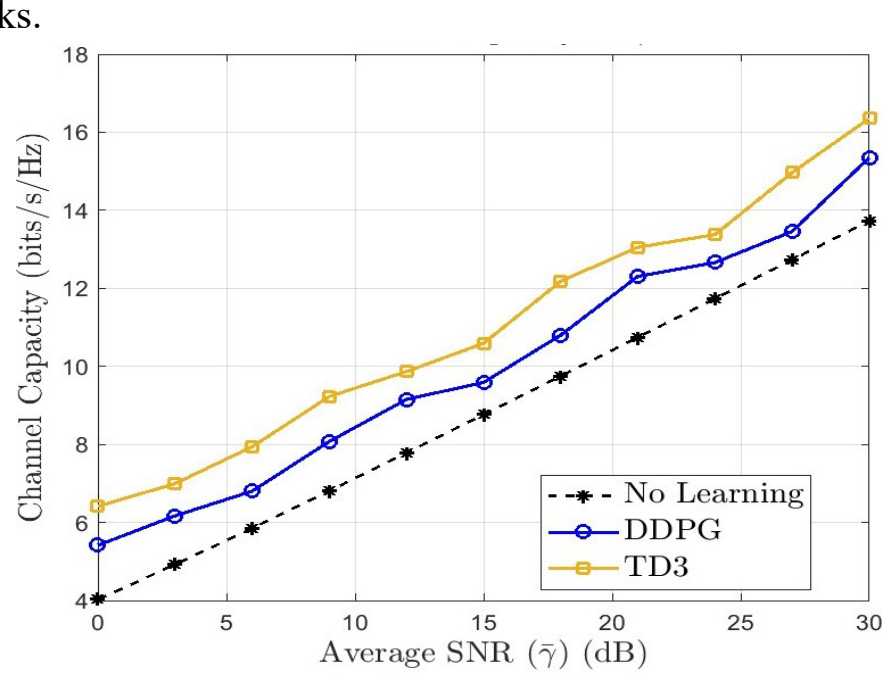

Fig. 8. Channel Capacity of the UWOC−O-RIS channel under DRL-based TD3, DDPG, and no learning.

## IV. Conclusion and Future work

This work presents a hybrid communication system that combines O-RIS-assisted UWOC with FSO/RIS-aided RF links under a hard-switching scheme. Using OTOPS-based channel modeling, the design takes into account turbulence and environmental effects. To handle the complexity of O-RIS phase optimization, a DRL-based control framework with DDPG and TD3 was introduced, aligning with IKT themes on intelligent optimization and AI-driven network control. Simulations show clear gains in outage probability, capacity, and reliability, with TD3 outperforming traditional methods. The framework can be extended in future work to cooperative multi-RIS scenarios with adaptive switching across RF, FSO, and UWOC links. Integrating advanced ISAC techniques may further improve situational awareness and spectrum efficiency in emerging 6G cross-domain networks.

## V. References

[1] S. A. H. Mohsan et al., “Recent advances, future trends, applications and challenges of Internet of Underwater Things (IoUT): A comprehensive review,” J. Mar. Sci. Eng., vol. 11, p. 124, 2023.

[2] Ramavath, P. N., & Chung, W. Y. “Performance evaluation of re-configurable intelligent surface-assisted underwater and free-space wireless optical communication in the skip-zones”. *ICT Express*, *10*(2), 320-329.(2024)

[3] S. Mondal et al., “Outage probability analysis of hard-switching based mixed FSO/IRS-aided RF communication,” Proc. Nat. Conf. Commun. (NCC), Guwahati, India, pp. 1–6, 2023.

[4] A. Heydaribeni and H. Beyranvand, “Performance analysis of underwa ter optical wireless communication using O-RIS and fiber optic backhaul (extended version),” 2025, arXiv:2508.18915.

[5] Y. Ata *et al.*, “A unified channel model for IRS-aided underwater OWC with combined attenuation losses,” *IEEE J. Sel. Areas Commun.*, vol. 43, no. 5, pp. 1552–1567, 2025.

[6] H. Zhou, M. Erol-Kantarci, Y. Liu and H. V. Poor, "A Survey on Model-Based, Heuristic, and Machine Learning Optimization Approaches in RIS-Aided Wireless Networks," in IEEE Communications Surveys & Tutorials, vol. 26, no. 2, pp. 781-823, Secondquarter 2024.

[7] R. Hashemi et al., Deep reinforcement learning for practical phase shift optimization in RIS-assisted networks over short packet communications, Joint EuCNC/6G Summit, 2022, pp. 518–523.

[8] R. Hashemi et al., "Deep RL for Phase-Shift Optimization in RIS-Aided MISO URLLC," IEEE IoT J., vol. 10, no. 10, pp. 8931–8943, May 2023.

[9] T. Van Chien et al., "Coverage Probability and Ergodic Capacity of IRS-Enhanced Systems," IEEE Commun. Lett., vol. 25, no. 1, pp. 69–73, Jan. 2021.